\documentclass{article}[12pt]
\usepackage{graphicx} 
\usepackage[letterpaper,left=1.25in,top=1.25in,bottom=1.25in,right=1.25in,includeheadfoot]{geometry}
\usepackage{quantikz}
\usepackage{mathtools}
\usepackage{amsmath}
\usepackage{url}
\usepackage{multirow}
\usepackage{authblk}

\title{Quantum Advantage in Trading: A Game-Theoretic Approach}

\author[1,2]{Faisal Shah Khan\thanks{faisal\_khan@kenan-flagler.unc.edu.}}
\author[3,4]{Norbert M. Linke}
\author[4]{Anton Trong Than}
\author[5]{Dror Baron}
\affil[1]{Taqtics LLC, Apex, NC 27539, USA }
\affil[2]{Rethinc. Labs, Kenan–Flagler Business School, University of North Carolina, Chapel Hill, NC 27599, USA}
\affil[3]{Duke Quantum Center and Department of Physics, Duke University, Durham, NC 27701, USA}
\affil[4]{Joint Quantum Institute and Department of Physics, University of Maryland, College Park, MD 20742, USA}
\affil[5]{Department of Electrical and Computer Engineering, North Carolina State University, Raleigh, NC 27606, USA}

\date{}

\begin{document}

\maketitle

\begin{abstract}
Quantum games, like quantum algorithms, exploit quantum entanglement to establish strong correlations between strategic player actions. This paper
introduces quantum game-theoretic models applied to trading and demonstrates their implementation on an ion-trap quantum computer. The results showcase a quantum advantage, previously known only theoretically, realized as higher-paying market Nash equilibria. This advantage could help uncover alpha in trading strategies, defined as excess returns compared to established benchmarks. These findings suggest that quantum computing could significantly influence the development of financial strategies.
\end{abstract}

\section{Introduction}

The nature of trading has consistently been shaped by advancements in information delivery and technology. From the introduction of the telegraph in the 1850s to the development of the stock ticker by Edward Calahan in 1867, real-time reporting of stock prices and market news revolutionized the trading landscape. These innovations allowed brokers and traders to access information more rapidly than ever before, fostering a more dynamic and competitive market environment. 

The digital era brought a paradigm shift with the advent of online trading platforms. The launch of digiTRADE in 1994 and Ameritrade’s pioneering online brokerage services enabled traders to place orders instantaneously with minimal or even no commissions. The internet introduced unprecedented speed and accessibility while offering tools like data integration, dashboards, and business intelligence for more informed decision-making. This transition to electronic trading also marked the decline of traditional floor traders and brokers, as algorithmic trading took center stage. Automated systems capable of analyzing market conditions and executing trades underscored the growing reliance on technology in the trading ecosystem.

Building on these developments, a key question involves the potential advantages of a quantum computing-based ``quantum trading'' platform. Current advancements in quantum computing suggest increased speed and efficiency in data processing \cite{Dong, Herman}; however, it is the proven ability of quantum computers to deliver higher-quality solutions to competitive interaction problems that offers unique value to traders \cite{Khan}. This can be particularly relevant for mission-critical markets such as carbon trading and other green markets. This paper examines game-theoretic trading models using the games Chicken and Prisoner's Dilemma as prominent examples to implement on a quantum trading platform.

\section{Core Trading Strategies: Long and Short}

Trading fundamentally revolves around the actions of buying and selling. When directed toward achieving specific objectives, these actions evolve into strategies designed to meet those goals. In this strategic context, buying is termed ``going long,'' and selling is referred to as ``going short.'' However, these terms are not merely catchy labels; their meaning and execution vary depending on the market in which the trading takes place.

For instance, in futures trading, going long entails purchasing a futures contract with the expectation that the price of the underlying asset—such as a commodity, stock index, or cryptocurrency—will rise. Conversely, going short involves selling a futures contract with the anticipation that the price of the underlying asset will fall.

In contrast, within the equity market, the focus of this discussion, going short involves borrowing shares of a company from a broker and selling them at the current market price, expecting a subsequent price drop due to increased supply. If the price falls as predicted, the trader repurchases the shares at the lower price, returns them to the broker (along with any applicable fees), and retains the difference as profit. Here, going long signifies buying first and selling later, while going short reverses the order: selling first and buying later. While short trading can aid in price discovery, its excessive use may contribute to financial crises like the one that occurred in 2008 \cite{Cruttenden}.

Regardless of the specific market, these trading actions hold strategic significance only when driven by an underlying ``game'' or competitive dynamic. Depending on the nature of this game, the outcomes of these strategies can vary significantly. For example, consider a simple scenario in the equity market where two traders—perhaps large fund managers—engage in a game of ``Chicken,'' facing off with payoffs that include not just monetary gains or losses but also intangible stakes like bragging rights or reputation. Let us formalize both the game Chicken and its manifestation in trading. 

\subsection{Trading as Chicken}

The classic narrative of the game Chicken involves two drivers speeding toward each other on a narrow road. At some point, they must decide how to navigate the impending encounter. The road is so narrow that both cannot pass simultaneously at full speed; at least one driver must slow down to avoid a collision. In this scenario, each driver faces two strategies: either to ``go slow''  or to ``speed up.'' The former represents a cautious, safety-first approach, but comes with the social cost of appearing to yield, or losing face. The latter strategy embodies taking a bold risk, with the potential reward of earning bragging rights if the other driver backs down. To emphasize the physics, and ultimately quantum physics, underlying game theory, the players will signal their strategic choice using coins: heads ($H$) for going slow and tails ($T$) for speeding up.  

\begin{figure}
    \centering
    \includegraphics[scale=0.55]{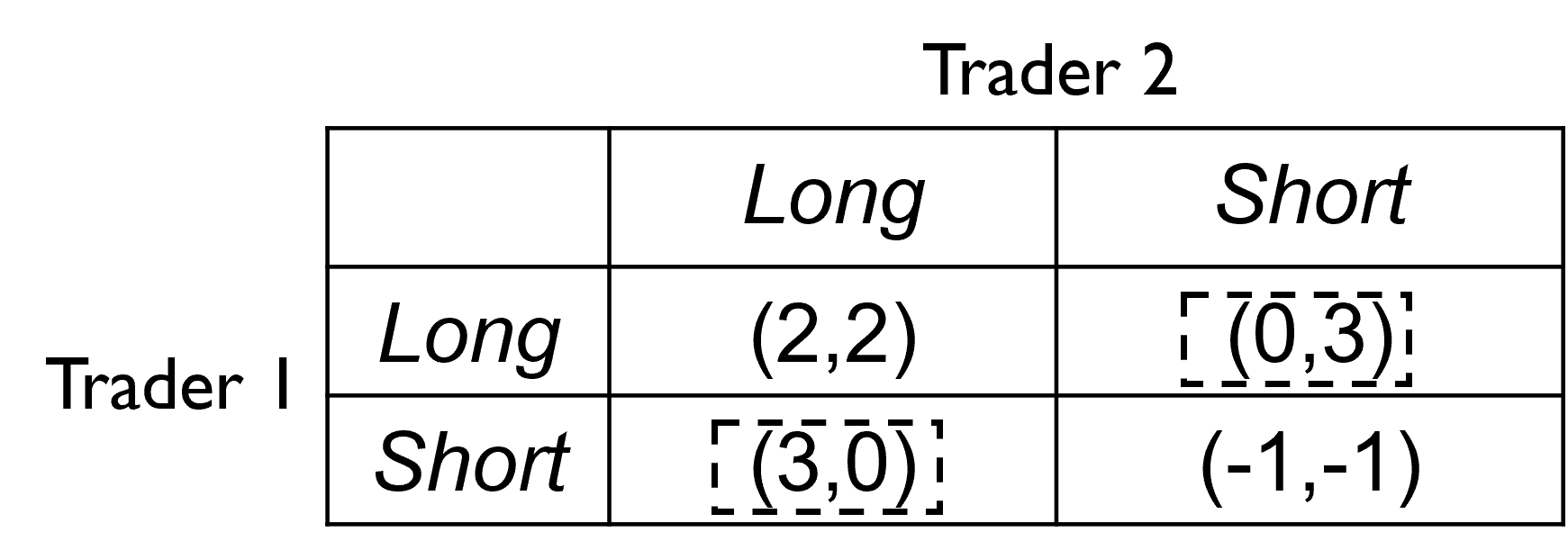} 
    \caption{Trading as the game Chicken with payoff determined by the strategies {\it Long} versus {\it Short}. The dashed outcomes, along with their corresponding strategy pairs, represent Nash equilibria. }
    \label{fig:Chickn}
\end{figure}

In the context of trading, these strategies can be reframed: going slow corresponds to going long, while speeding up aligns with going short. To clarify the dynamics of this interaction, we can assign specific payoffs to the players based on the strategies they choose, creating a payoff matrix that reflects the outcomes of the game. The matrix shown in Figure \ref{fig:Chickn}, as taken from \cite{binmore}, represents the strategic options in this scenario, {\it Long} or {\it Short}, and the resulting payoff to the players. The first number in a tuple is the payoff to the row player (Trader 1) and the second number is the payoff to the column player (Trader 2). 

A fundamental assumption in game theory is that all players are {\it rational}, meaning they consistently act to maximize only their individual payoffs. Consequently, a key solution concept in understanding competitive interactions is the Nash equilibrium: a collection of strategies—one for each player—where each strategy is a best response to all others. Put differently, it is an outcome where no player would benefit from changing their strategy unilaterally. 

In the game Chicken, there are two Nash equilibria, $(Long, Short)$ and $(Short,Long)$. Additionally, there exists a ``mixed'' strategy Nash equilibrium, which arises when players randomize between their original ``pure'' strategies. This randomization can be affected by the players tossing their coins and deciding to play the strategy corresponding to the resulting face. For mixed strategy equilibrium to exist, the probabilities determined by the coin toss should make the other player indifferent about which strategy to choose as a best response. In Chicken, the mixed strategy Nash equilibrium is achieved when both players choose \textit{Long} 50\% of the time and \textit{Short} 50\% of the time. This results in an expected payoff of $1$ for each player. Although this payoff is lower than the pure strategy payoff of $3$, the advantage of the mixed strategy Nash equilibrium is that it ensures no player receives a payoff of $0$, offering a measure of risk mitigation in the game. 

\subsection{Trading as Prisoner's Dilemma}

\begin{figure}
    \centering
    \includegraphics[scale=0.55]{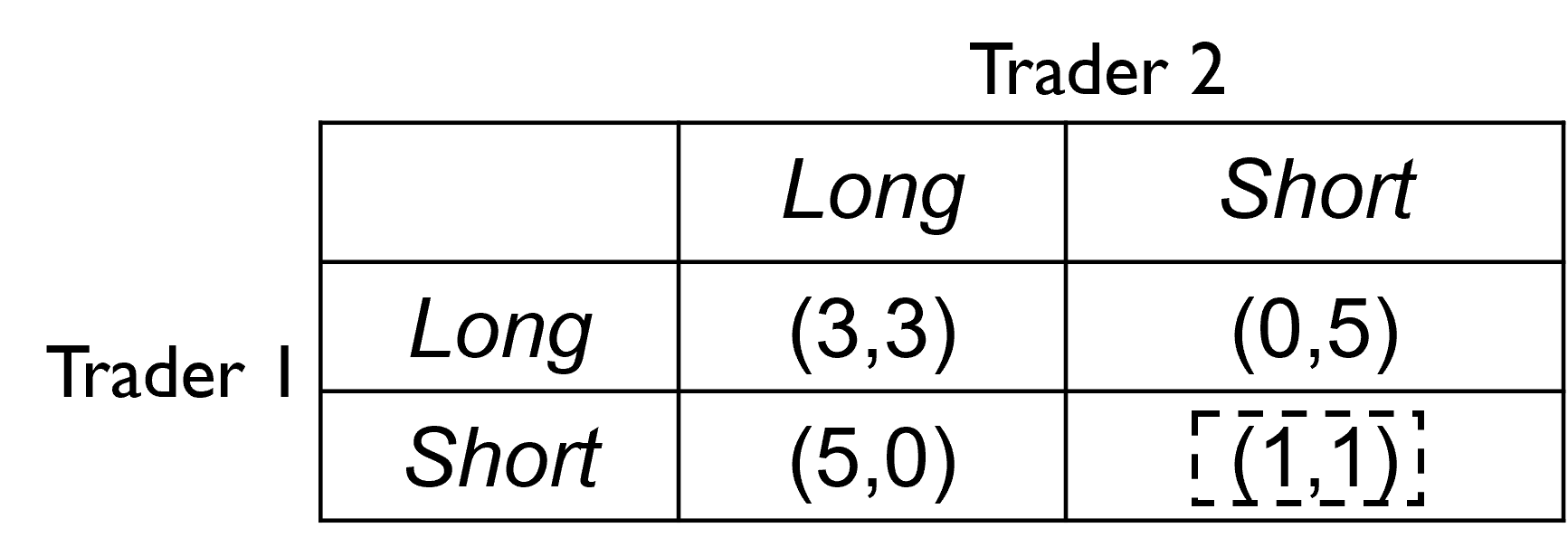} 
    \caption{Trading as the game Prisoner's Dilemma with payoffs determined by the strategies {\it Long} versus {\it Short}. The unique Nash equilibrium is at the dashed outcome $(1,1)$ corresponding to the strategy pair (\textit{Short, Short}). }
    \label{fig:PD}
\end{figure}

Prisoner's Dilemma illustrates how individually rational strategies can lead to outcomes that are suboptimal for the group. This scenario is reflected in a trading context, as shown in Figure \ref{fig:PD} (also sourced from \cite{binmore}), where the strategy of {\it Short} consistently offers a higher expected payoff for each trader: $5$ or $1$, compared to $3$ or $0$ when playing {\it Long}.

Imagine a two-trader market snapshot, where both traders are seeking to profit from equity market investments. If both choose to go long, the market remains relatively stable, and each trader can expect a steady profit of $3$. However, the temptation to go short is strong: if one trader shorts while the other goes long, the shorting trader can secure a substantial profit of $5$, while the long trader makes no profit.

Shorting, however, comes with significant risk. If both traders decide to play {\it Short}, their profits are limited to $1$ each due to the heightened risk of a short squeeze—a situation where rising market prices force short sellers to cover their positions, driving prices even higher. However, when these rational choices are played out, the resulting Nash equilibrium is exactly $(Short, Short)$ where each trader only gets a payoff of $1$.

\section{Refereed Trading}
In his Nobel Prize winning work \cite{Aumann}, Aumann showed how it is possible to enhance the quality of Nash equilibria by introducing correlation into the players' strategic actions. For example, in the game of Chicken, traders may correlate their actions by connecting their coins with a flexible wire or cable, creating a system we call a referee. 
This referee establishes a publicly known probability distribution over the four possible outcomes of the game, determined by the properties of the coins and the connecting wire.
\begin{figure}
    \centering
    \includegraphics[scale=0.55]{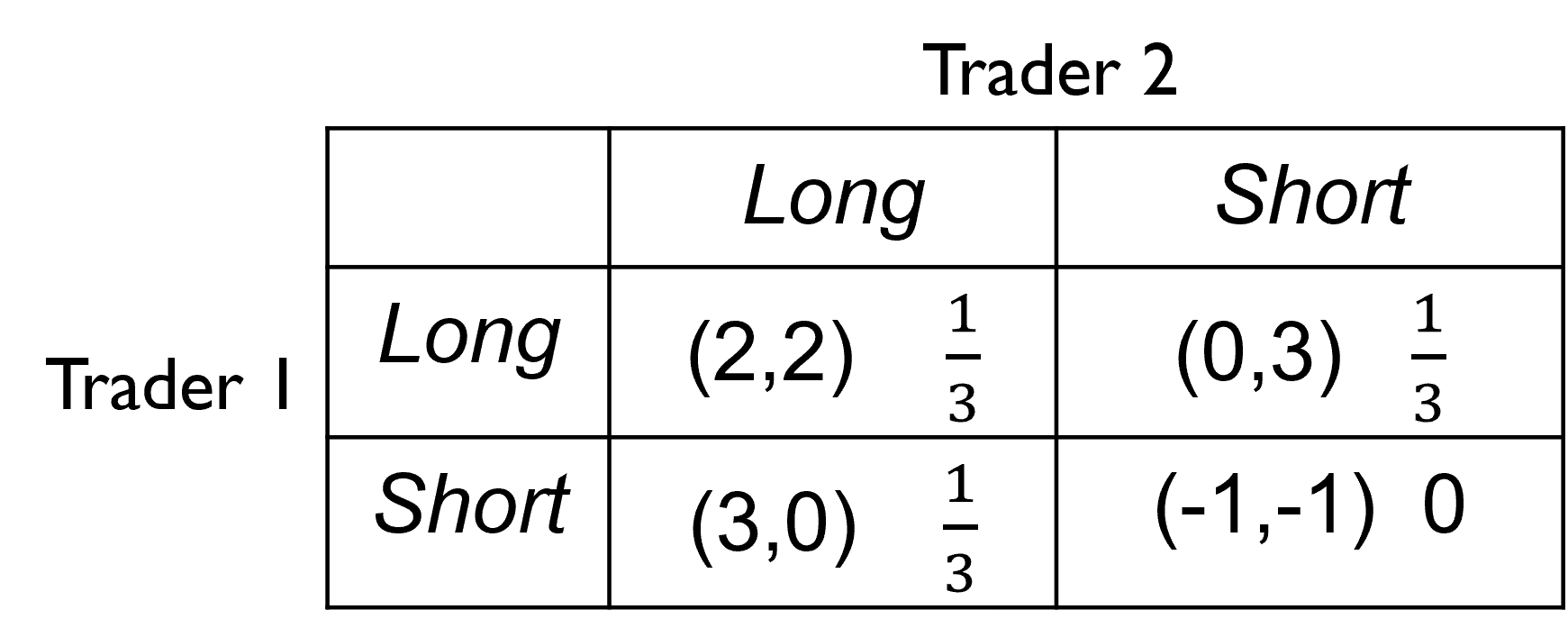} 
    \caption{Trading as the game Chicken with a referee characterized by the probability distribution $\left( \frac{1}{3}, \frac{1}{3}, \frac{1}{3},0 \right)$.  }
    \label{fig:refChickn}
\end{figure}
After tossing their individual coins, each player observes his own outcome. However, because of the wire connection, these outcomes are no longer independent and the player also has to take into account the correlation between the coin tosses. This is done by assessing the probabilities with which his opponent will play her strategies, conditional on the result of his coin toss. 

This process can be conceptualized as the referee providing advice to each player on what strategy to play (the individual coin toss outcome). The players then decide whether to follow this advice by predicting what advice was given to their opponent. This prediction process involves computing conditional probabilities, where the conditioning event is a player's own coin toss outcome, and evaluating the expected payoffs of agreeing or disagreeing with the referee's suggestion. When both players always follow the referee's advice, the resulting outcome is referred to as a {\it correlated equilibrium}. 

In the game of Chicken, for example, if the referee is characterized by the probability distribution $(\frac{1}{3},\frac{1}{3},\frac{1}{3},0 )$ over the game’s outcomes as given in Figure \ref{fig:refChickn}, both players will always agree with the referee's advice and earn an expected payoff of $\frac{5}{3}$ each, an improvement over the mixed strategy Nash equilibrium. To understand how this improvement arises, consider the case where Trader 2 tosses her coin and receives advice to play \textit{Long}. Let this event be denoted as $B$. To decide whether to follow the advice or deviate from it, Trader 2 evaluates the following scenarios:
\begin{itemize}
    \item $A_1$: Trader 1 is advised by the referee to play \textit{Long}.
\item $A_2$: Trader 1 is advised by the referee to play \textit{Short}.
\end{itemize}
The conditional probabilities of these events, given $B$, are calculated as:
\begin{equation}
        P(A_1|B) = P(A_2|B) =\frac{\frac{1}{3}}{\frac{1}{3}+\frac{1}{3}}=\frac{1}{2}.
\end{equation}

The expected payoff to Trader 2 from agreeing with the referee is therefore $2 \cdot \frac{1}{2}+0 \cdot \frac{1}{2}=1$, and disagreeing gives her the payoff  $3 \cdot \frac{1}{2}+ (-1) \cdot \frac{1}{2}=1$. Since the payoffs are the same, Trader 2 is indifferent between the two options and hence agrees with the referee's advice and plays \textit{Long}. Her overall expected payoff in the game is $2 \cdot \frac{1}{3}+ 0 \cdot \frac{1}{3} + 3 \cdot \frac{1}{3} + (-1)\cdot 0= \frac{5}{3}$. Similar reasoning shows that both players will comply with the referee's advice if it is to play any strategy that is consistent with the probability distribution $(\frac{1}{3},\frac{1}{3},\frac{1}{3},0 )$. Consequently, this distribution constitutes a correlated equilibrium in the game of Chicken.

On the other hand, unlike Chicken, the Prisoner's Dilemma has the notable characteristic that the strategy {\it Short} strongly dominates {\it Long}. Hence, neither mixed strategies nor the introduction of a referee to correlate players' actions can lead to an improved Nash equilibrium. Regardless of the probability distribution over the game's outcome that characterizes the referee, both players will always deviate from the referee's advice to play {\it Long}. This rigid structure of the game's outcomes is part of what makes it both compelling and widely discussed.

The Prisoner's Dilemma is also a good case to illustrate how quantum computing can reshape strategic interactions. By employing quantum physical principles such as superposition and entanglement, quantum computers can bypass the dilemma's limitations, offering superior Nash equilibria that are otherwise unattainable. More dramatically, if quantum computing is available to only one player, it can create a clear advantage by enabling the quantum trader to achieve outcomes beyond the reach of their classical counterpart. 

\section{Quantum-Refereed Trading}

\begin{figure}
    \centering
        \begin{quantikz}
\ket{0} \quad &\gate[2]{J}&\gate{U_1}&\gate[2]{J^{\dag}} & \\
\ket{0} \quad & & \gate{U_2}& & 
\end{quantikz}
    \caption{The EWL game quantization protocol for two-player, two-strategy games. The gate $J$ puts the two qubits $\ket{00}$ into the maximally entangled state $\frac{1}{\sqrt{2}}\ket{00} - \frac{i}{\sqrt{2}}\ket{11}$, similar to how a referee correlated the strategies of players in a classical game setting. The single qubit gates $U_1$ and $U_2$ are the quantum strategies of the players.}
    \label{fig:EWL} 
\end{figure}
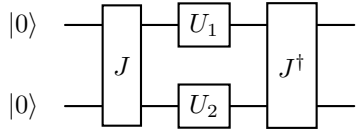

Mathematically, incorporating randomization and correlation through mixed strategies and a referee serves to extend the domain and range of the underlying game function. This process can also be applied in the quantum realm, enabling the inclusion of higher-order randomization and correlation. For a detailed exploration of this mathematical framework, refer to \cite{Bleiler}.

Eisert, Wilkens, and Lewenstein (EWL)~\cite{EWL1} introduced one such game extension in the form of a quantum circuit for two-player, two-strategy games. This EWL protocol is illustrated in Figure \ref{fig:EWL}. In this framework, players' coins are replaced with qubits, and correlations between classical coins are extended to maximal entanglement between qubits. Instead of probability distribution over game outcomes, the quantum referee is characterized by a quantum superposition of outcomes.

A key component of this protocol is the gate $J$, which generates entanglement between qubits in the following form:
\begin{equation}
\frac{1}{\sqrt{2}}\ket{00} - \frac{i}{\sqrt{2}}\ket{11}.
\end{equation}
This entanglement is reversed by $J^{\dag}$, the inverse of $J$. Mathematically, $J$ is a unitary matrix, meaning that applying the conjugate transpose operation, denoted by the symbol $\dag$, produces its inverses, $J^{\dag}$. The use of $J$ and $J^{\dag}$ ensures that the quantum model can reproduce the original game dynamics, accommodating both pure and mixed strategies. In the quantum setting, players implement their strategies by performing quantum operations—represented as gates $U_1$ and $U_2$—on their qubits. These operations are referred to as {\it quantum strategies}. The initial state of each qubit is $\ket{0}$, analogous to the heads ($H$) state of a coin, indicating the strategic choice \textit{Long}, while $\ket{1}$ corresponds to tails ($T$), indicating the strategic choice \textit{Short}.

The details of operation of the EWL quantum referee are as follows. The entangling and disentangling gates are, respectively,
\begin{equation}
    J:=e^{-i\frac{\pi}{4}(D \otimes D)}=\begin{pmatrix}
    \frac{1}{\sqrt{2}}& 0 & 0 & -\frac{i}{\sqrt{2}} \\
 0 & \frac{1}{\sqrt{2}} &  \frac{i}{\sqrt{2}} & 0 \\
 0 & \frac{i}{\sqrt{2}} &  \frac{1}{\sqrt{2}} & 0 \\
 - \frac{i}{\sqrt{2}}& 0 & 0 & \frac{1}{\sqrt{2}}    \end{pmatrix},
  \quad    J^{\dag}:=\begin{pmatrix}  \frac{1}{\sqrt{2}}& 0 & 0 & \frac{i}{\sqrt{2}} \\
 0 & \frac{1}{\sqrt{2}} &  -\frac{i}{\sqrt{2}} & 0 \\
 0 & -\frac{i}{\sqrt{2}} &  \frac{1}{\sqrt{2}} & 0 \\
 \frac{i}{\sqrt{2}}& 0 & 0 & \frac{1}{\sqrt{2}}
    \end{pmatrix},
\end{equation}
where $D$ is the quantum gate representing the original pure strategy \textit{Short}, defined as the unitary matrix
\begin{equation}
    \textit{Short} = D :=\begin{pmatrix}
   0 & 1\\
 -1 &  0
        \end{pmatrix}.
\end{equation}
The symbol $\otimes$ represents the tensor product of matrices. It is worthwhile to note here that the matrix representation of \textit{Short} is in fact the ubiquitous Pauli-Y matrix multiplied with the complex number $i$:
\begin{equation}
    i \cdot \begin{pmatrix}
   0 & -i \\
 i &  0
        \end{pmatrix}.
\end{equation}
For more details on quantum gates and their unitary matrix representations and operations, see \cite{Nielson}. 

The pure strategy \textit{Long} is represented as 
\begin{equation}
    \textit{Long} := \begin{pmatrix}
   1 & 0\\
 0 &  1
        \end{pmatrix}.
\end{equation}
In general, quantum strategies for player $k$, $k =1, 2$, are defined as unitary matrices 
\begin{equation}\label{qstrat}
 U(\theta_k, \phi_k) :=   \begin{pmatrix}
     e^{i\phi_k}\cos \frac{\theta_k}{2} &  \sin\frac{\theta_k}{2}\\
 - \sin\frac{\theta_k}{2} &  e^{-i\phi_k}\cos \frac{\theta_k}{2} 
        \end{pmatrix}, \quad 0 \leq \theta_k \leq \pi, \hspace{2mm} 0 \leq \phi_k \leq \frac{\pi}{2}.
\end{equation}
Note that \textit{Long} occurs when $\theta = \phi = 0$, while \textit{Short} occurs when $\theta = \pi$ and $\phi = 0$. The EWL quantum referee is then characterized by the quantum superposition $(\mu_1,\mu_2,\mu_3,\mu_4)$ of the outcomes $\left\{ \ket{00}, \ket{01}, \ket{10}, \ket{11}\right\}$ of the game, where
\begin{equation} \label{quantsup2}
\begin{split}
\mu_1 & :=\cos (\phi_1 + \phi_2) \cos \left( \frac{\theta_1}{2} \right) \cos \left(\frac{\theta_2}{2}\right),  \\
\mu_2 &:= -i\left[ \sin(\phi_2) \sin \left(\frac{\theta_1}{2} \right)\cos\left(\frac{\theta_2}{2} \right) - \cos(\phi_1) \cos\left(\frac{\theta_1}{2} \right) \sin\left(\frac{\theta_2}{2} \right) \right],\\
\mu_3 &:= -i\left[ \sin(\phi_1) \cos \left(\frac{\theta_1}{2} \right)\sin\left(\frac{\theta_2}{2} \right) - \cos(\phi_2) \sin\left(\frac{\theta_1}{2} \right) \cos\left(\frac{\theta_2}{2} \right) \right],\\
\mu_4 &:= \sin(\phi_1 + \phi_2) \cos\left(\frac{\theta_1}{2} \right)\cos\left(\frac{\theta_2}{2}\right) + \sin\left(\frac{\theta_1}{2} \right) \sin\left(\frac{\theta_2}{2} \right).
\end{split}
\end{equation}

The quantum superposition (\ref{quantsup2}) is analogous to the probability distribution that characterizes a classical referee. Indeed, when measured, it gives a probability distribution over the outcomes of the game. Similarly, the quantum strategies can be viewed as the players' actions of ``tossing'' their respective qubits within a maximally entangled two-qubit system. These tosses place the qubits into a quantum superposition of $\ket{0}$ and $\ket{1}$, representing the quantum referee's advice, with which the players may agree or disagree. If both players agree, then the result is a quantum correlated equilibrium, a further refinement of Nash equilibrium. 

\subsection{Quantum Strategies and Alpha}

For the Prisoner's Dilemma, when the quantum referee advises both players to employ the quantum strategy 
\begin{equation}\label{qlong}
 \textit{quantum Long} :=    \begin{pmatrix}
    i & 0\\
 0 &  -i
        \end{pmatrix},
    \end{equation}
with $\theta=0, \phi=\frac{\pi}{2}$, they will comply. It can be shown mathematically and verified experimentally that deviating from the referee's advice of {\it quantum Long} and playing any other quantum strategy from the set in definition (\ref{qstrat}) yields a smaller payoff. Therefore, {\it quantum Long} is a best reply to itself, resulting in the quantum correlated (Nash) equilibrium with corresponding probability distribution $(1,0,0,0)$ over the outcomes so that each player receives a payoff of $3$. 

In Chicken, the same dynamics give both traders a payoff of $2.$ More specifically, if one trader plays the conventionally dominant strategy \textit{Short} against {\it quantum Long}, his payoff is $0$. This demonstrates the advantage of quantum refereed trading: a trader unaware that the trading environment has shifted to quantum computation will consistently receive a payoff of $0$ against the quantum trader who will generate alpha due to his superior strategy. 

The correlations that characterize the EWL quantum referee differ fundamentally from those achievable in conventional settings. To illustrate this, consider when Trader 1 is advised by the quantum referee to go long. Assuming a conventional mindset though, Trader 1 deviates from this advice by flipping over his qubit to indicate that he is going short. To his surprise, the qubit remains in the state $\ket{0}$. Even more surprising is that his action flips the qubit of Trader 2, leaving it in the state $\ket{1}$. This is impossible when trades are made under the advice of a conventional, {\it honest} referee characterized by coins, especially if the coins were located a great distance apart \cite{Clauser}. 

The two-player EWL quantum referee has been experimentally implemented, as demonstrated by Solmeyer et al. \cite{Solmeyer}. Additionally, alternative quantum referee models have been investigated, including the one introduced by Chappell et al. \cite{Chap}, which was explored within the framework of the Einstein-Podolsky-Rosen (EPR) experiment.

\subsection{Mixed Quantum Strategies and Alpha}

When the set of quantum strategies is expanded to include a full class of gates parametrized by three real parameters, for example, 
\begin{equation}\label{fullqstrat}
U(\alpha_k, \theta_k, \gamma_k) :=  
\begin{pmatrix}
e^{i\alpha_k} \cos\frac{\theta_k}{2} & e^{i\gamma_k} \sin\frac{\theta_k}{2} \\
-e^{-i\gamma_k} \sin\frac{\theta_k}{2} & e^{-i\alpha_k} \cos\frac{\theta_k}{2}
\end{pmatrix}, \quad \alpha_k, \gamma_k \in [0,2\pi], \theta_k \in [0, \pi],
\end{equation}
the quantum alpha previously obtained via quantum strategies vanishes. This occurs because, in this larger space, each player can devise a counter-strategy to any quantum strategy employed by the opponent \cite{EWL2}.

However, an additional advantage emerges when the players employ their quantum strategies probabilistically, using {\it mixed} quantum strategies. In this approach, a player tosses her qubit one way a certain percentage of the time and a different way the remaining percentage of the time. This results in two distinct quantum superpositions of $\ket{0}$ and $\ket{1}$, which can be interpreted as the quantum referee providing probabilistic advice to the players.

For instance, suppose the referee advises Trader 1 to play \textit{Long} half of the time and the quantum strategy 
\begin{equation} 
    \textit{quantum Long \#1}:=    \begin{pmatrix}
    -i & 0\\
 0 &  i
        \end{pmatrix}
\end{equation}
the other half of the time. Similarly, suppose Trader 2 is advised to play \textit{Short} half the time and the quantum strategy
\begin{equation} 
    \textit{quantum Short}:=    \begin{pmatrix}
    0 & -i\\
 -i &  0
        \end{pmatrix}
\end{equation}
the other half of the time. The resulting quantum superpositions yield the probability distributions $(0,1,0,0)$ and $(0,0,1,0)$ over the outcomes, each occurring with equal probability. These combine to form the effective distribution $\left( 0, \frac{1}{2}, \frac{1}{2},0  \right)$. These mixed quantum strategies are best responses to each other and thus form a \textit{mixed quantum correlated equilibrium}, where each trader receives a payoff of $2.5$. This is slightly less than the ideal payoff of $3$, but still represents a significant improvement over the original payoff of $1$.

\section{Multiple Traders}

\begin{figure}
    \centering
     \centering
    \includegraphics[scale=0.55]{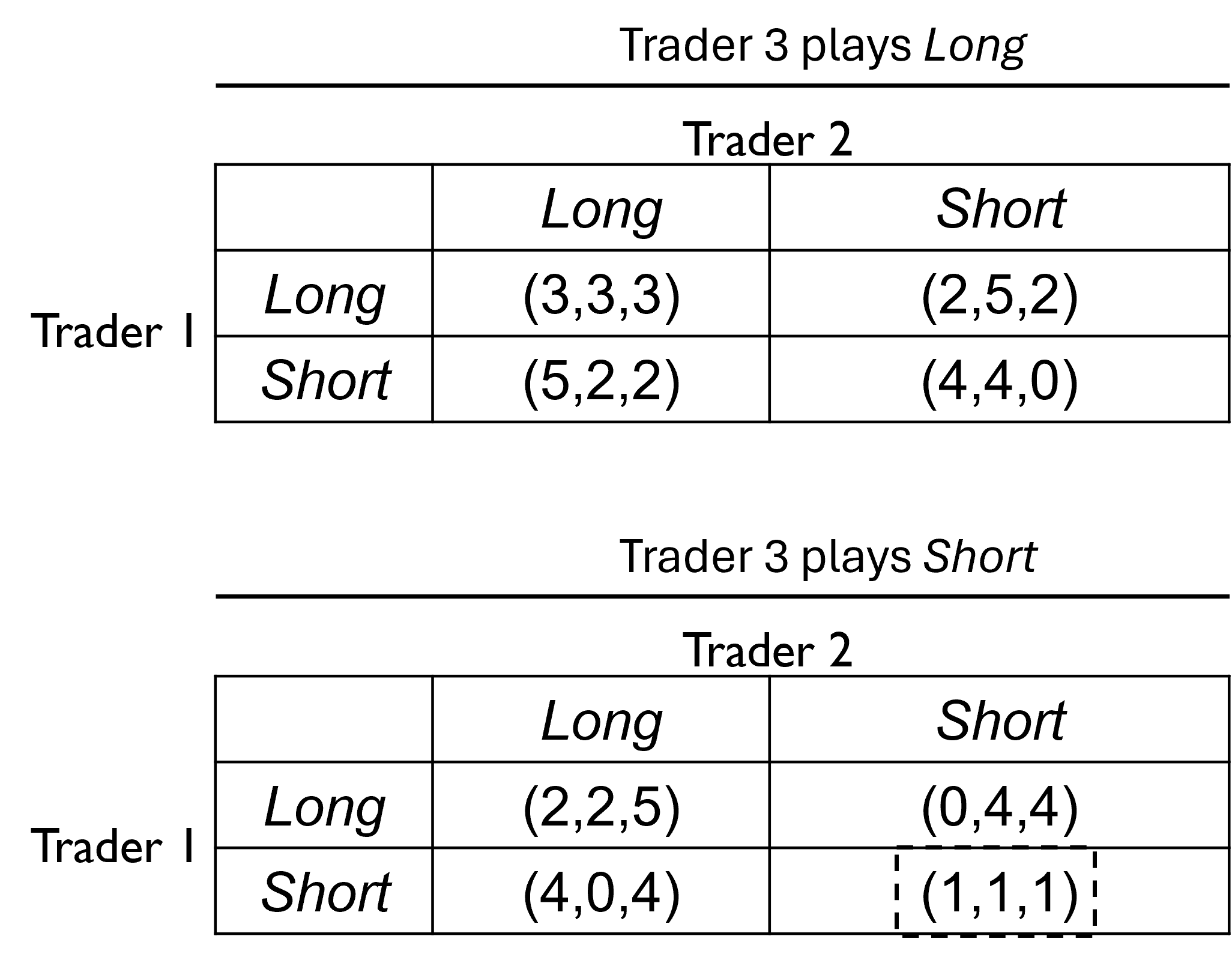} 
    \caption{Three player Prisoner's Dilemma model of with a unique, suboptimal Nash equilibrium (\textit{Short, Short, Short}), highlighted in a dashed box, that cannot be improved upon with mixing or the introduction of a referee.}
    \label{fig:PDmult}
\end{figure}

\begin{figure}
    \centering
    \includegraphics[scale=.55]{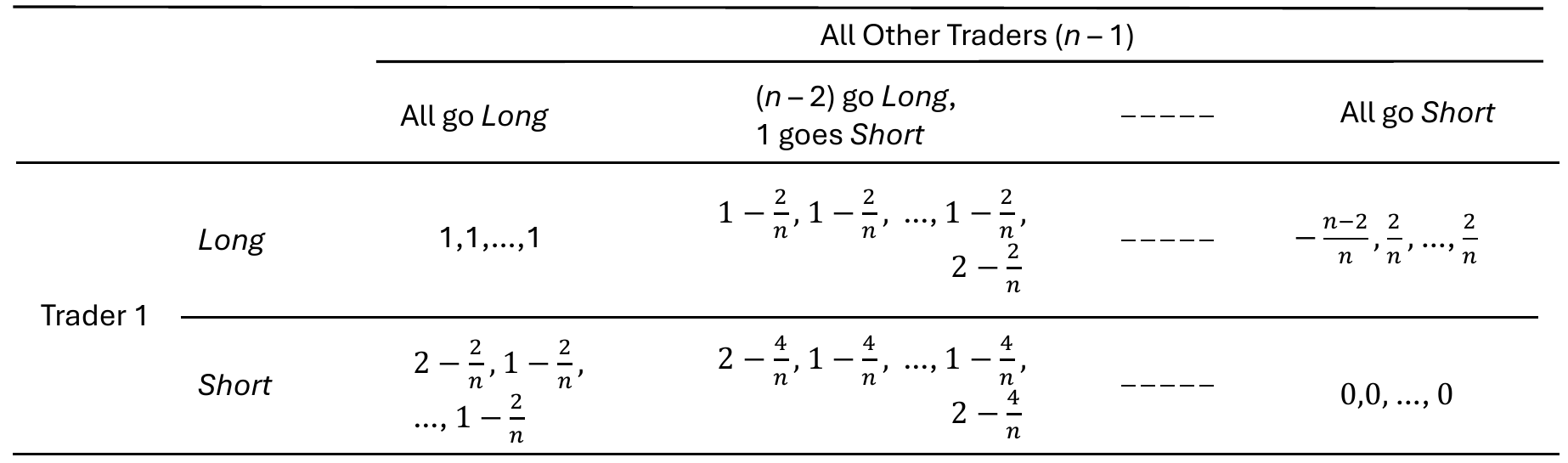}
    \caption{A $n$-trader Prisoner's Dilemma model for trading. }
    \label{fig:nplayerPD}
\end{figure}

The game-theoretic models of equity trading and their quantum-refereed implementations can be extended to include multiple players. For three-trader Prisoner's Dilemma, the version depicted in Figure \ref{fig:PDmult} is often used in quantum game theory literature. A best-response analysis of this version reveals that the only pure strategy Nash equilibrium is (\textit{Short, Short, Short}), highlighted with a dashed box, where each trader only gets $1$. A generalized version for $n$-trader Prisoner's Dilemma is documented in \cite{Magli} and is replicated in Figure \ref{fig:nplayerPD}. We use this version in our quantum implementation of trading for $n=3,4,5$, and $6$. Note that in this version of the Prisoner's Dilemma, the Nash equilibrium where everyone goes \textit{Short} produces a payoff of $0$, while a payoff of $1$ is made if all traders were to go \textit{Long}.

Even with the introduction of mixed strategies, the traders' payoffs do not improve. Furthermore, employing a referee to correlate the players' actions fails to enhance the outcomes, regardless of the probability distribution used to characterize the referee. To illustrate, consider a referee modeled by three coins interconnected via some mechanism (e.g., a wire or cable). The referee's influence is represented by a probability distribution $(p_1, p_2, \cdots, p_8)$ over the eight possible outcomes of the game.

Assume that Trader 3 tosses her coin and receives advice to play \textit{Long}. We denote this event as $X$. To decide whether to follow this advice or instead play \textit{Short}, Trader 3 considers the following scenarios:
\begin{itemize}
    \item $A_1$: Both Trader 1 and Trader 2 are advised by the referee to play \textit{Long}.
\item $A_2$: Trader 1 is advised to play \textit{Short}, and Trader 2 to play \textit{Long}.
\item $A_3$: Trader 1 is advised to play \textit{Long}, and Trader 2 to play \textit{Short}.
\item $A_4$: Both Trader 1 and Trader 2 are advised to play \textit{Short}.
\end{itemize}
The conditional probabilities of these events, given $X$, are calculated as:
\begin{equation}
        P(A_i|X) =\frac{p_i}{p_1+p_2+p_2+p_4}; \quad i=1,2,3,4.
\end{equation}

Trader 3 then evaluates her expected payoff for two options: agreeing with the referee's advice or disregarding it. Using the version of Prisoner's Dilemma in Figure \ref{fig:PDmult}, the expected payoff for agreeing is:
\begin{equation}\label{agree}
  3 \cdot P(A_1|X)+2 \cdot P(A_2|X)+2 \cdot  P(A_3|X)+0 \cdot P(A_4|X).
  \end{equation}
On the other hand, the expected payoff for disagreeing and playing \textit{Short} is:
\begin{equation}\label{disagree}
    5\cdot P(A_1|X)+4 \cdot  P(A_2|X)+4 \cdot P(A_3|X)+ 1 \cdot P(A_4|X).
\end{equation}
By comparing these equations, it becomes clear that Trader 3 will always achieve a higher expected payoff by disagreeing with the referee and playing \textit{Short}. The three-trader Prisoner's Dilemma also demonstrates the strong dominance of the \textit{Short} strategy over \textit{Long}. A similar analysis will show that this result holds for the version of Prisoner's Dilemma in Figure \ref{fig:nplayerPD}.

\subsection{Quantum Referee: Three Traders and Beyond} \label{sec:q3traders}

In \cite{Du}, Du et al. propose an extension of the EWL quantum referee to three or more players (traders) using the payoff table in Figure \ref{fig:PDmult}. These author use as the entangling gate
\begin{equation}
    \mathcal{J}:=e^{i\frac{\pi}{4}(\sigma_x \otimes \sigma_x \otimes \sigma_x)}
\end{equation}
where 
\begin{equation}
    \sigma_x:=  \begin{pmatrix}
 0 & 1\\
 1 & 0
        \end{pmatrix},
\end{equation}
is the unitary matrix for the quantum Pauli-X gate. Their quantum strategies are parametrized as
\begin{equation}\label{qstratDU}
 U(\theta_k, \phi_k) :=   \begin{pmatrix}
    \cos \frac{\theta_k}{2} &  e^{i\phi}\sin\frac{\theta_k}{2}\\
 -e^{-i\phi} \sin\frac{\theta_k}{2} &  \cos \frac{\theta_k}{2} 
        \end{pmatrix}
\end{equation}
with the parameters $\theta_k$ and $\varphi_k$ taking values in the same range as in the EWL parametrization. 

For the case of three traders, the quantum referee is characterized by a quantum superposition of the outcomes of the underlying game that generalizes the quantum supposition in (\ref{quantsup2}). This generalization is the result of adding a qubit, in the state $\ket{0}$, to the quantum referee mechanism together with quantum strategies of the form in (\ref{qstratDU}) for Trader 3. In \cite{Dax}, the authors give an analytic expression for the quantum superposition characterizing the EWL quantum referee in an $n$-player scenario. 

The quantum entanglement between the three qubits is given by the phase-adjusted Greenberger-Horne-Zeilinger (GHZ) state
\begin{equation}
    \frac{1}{\sqrt{2}}\ket{000}+ \frac{i}{\sqrt{2}}\ket{111}.
\end{equation}
Du et al.'s quantization for three players reaches a quantum correlated equilibrium when the referee advises the traders to play \textit{Short}. However, this equilibrium results in a payoff of 3 for each trader! 
The study of mixed quantum correlated equilibria in games involving three or more traders remains an under explored area. Notable progress in this direction is the work of Ahmed in \cite{Ahmed}.

\section{Implementation on Ion-Trap Quantum Computer}

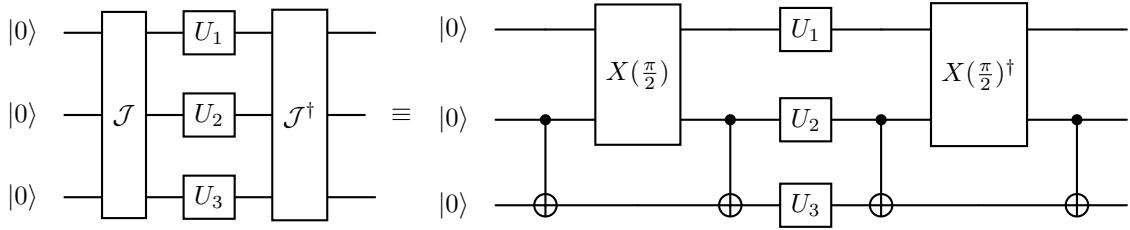
\begin{figure}
    \centering
        \begin{quantikz}
\ket{0} \quad & \gate[3]{\mathcal{J}} &\gate{U_1} & \gate[3]{\mathcal{J}^{\dag}} & \quad  \\
\ket{0} \quad & & \gate{U_2} & &  \quad  \equiv \\
\ket{0} \quad & & \gate{U_3} & & \quad 
\end{quantikz}
  \begin{quantikz}
\ket{0} \quad &  & \gate[2]{X(\frac{\pi}{2})} &  & \gate{U_1} & & \gate[2]{X(\frac{\pi}{2})^{\dag}}  & & \\
\ket{0} \quad & \ctrl{1} & & \ctrl{1} &   \gate{U_2} & \ctrl{1} & & \ctrl{1}& \\
\ket{0} \quad & \targ{} &  &  \targ{} &   \gate{U_3} & \targ{} & &  \targ{} &
\end{quantikz}
    \caption{The three-trader, two-strategy game quantization (quantum referee) a la Du et al. The gates $\mathcal{J}$ and its inverse in the diagram on the left decompose into two controlled-not (CNOT) gates and one entangling gate $X(\frac{\pi}{2})$ to create the GHZ state of the three qubits, each initialized to $\ket{0}$. The single qubit gates  are the quantum strategies of the players.}
    \label{fig:EWL3} 
\end{figure}

Building on the work of Du et al., we implemented the quantum-refereed Prisoner's Dilemma trading model for three or more traders on an ion-trap quantum computer housed at the University of Maryland \cite{Debnath}. The system consists of a linear chain of trapped $^{171}\text{Yb}^+$ ions; qubits are encoded in the energy levels the ion.  During operation, the qubits are initialized to the $\ket0$ state through optical pumping, and quantum gates are performed by turning on laser beams that individually address each ion.  Measurement is then done through state-dependent fluorescence \cite{Olmschenk}.

\begin{figure}
    \centering
    \includegraphics[width=0.65\textwidth]{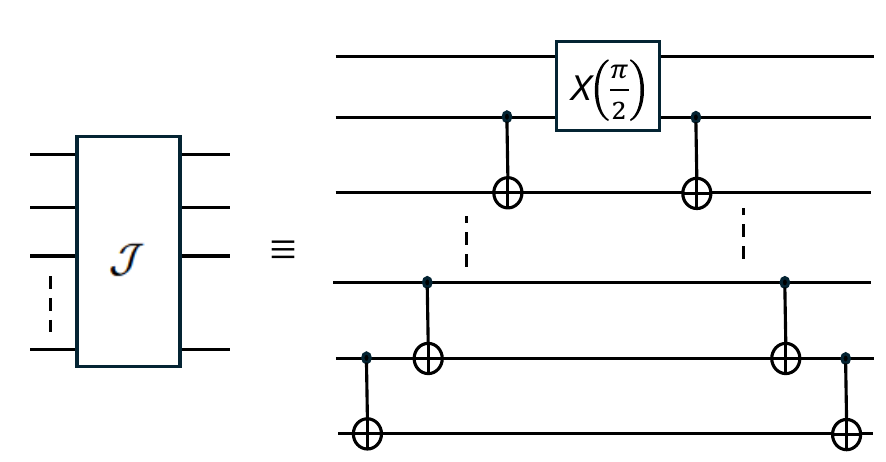}
    \caption{Decomposition of $n$ qubit entangling operation $\mathcal{J}$ in terms of CNOT gates and the two qubit entangling gate $X(\frac{\pi}{2})$.}
    \label{fig:entangling_gate}
\end{figure}

The native operations of the system include a universal gate set of single-qubit rotations and an entangling two-qubit gate $X(\theta) = e^{-i\theta\sigma_x\otimes\sigma_x}$ with all-to-all connectivity, mediated by a M{\o}lmer-S{\o}rensen interaction \cite{MS}. Each entangling gate takes around 300 $\mu$s with a fidelity of approximately 98.5\%.  For each protocol, the algorithm was compiled onto the native gate-set, and then simplified to minimize the number of entangling gates in each circuit.
The construction of $\mathcal{J}$ for three or more qubits utilizes the two-qubit entangling and disentangling gates $X(\frac{\pi}{2})$ and $X(\frac{\pi}{2})^{\dag}$, along with CNOT gates, as shown in Figures \ref{fig:EWL3} and \ref{fig:entangling_gate}. 

To experimentally demonstrate the Nash equilibrium,  we allow Trader 1 to vary his quantum strategy parameters \((\theta, \phi)\) while holding the strategies of the other \(n-1\) traders fixed at the equilibrium parameters \((\pi, 0)\). The payoff to Trader 1 is then measured. By symmetry, if the payoff is maximized when Trader 1 adopts the parameters \((\pi, 0)\), this serves as evidence that the Nash equilibrium has been achieved.

\begin{figure}
    \centering
    \includegraphics[width=0.8\textwidth]{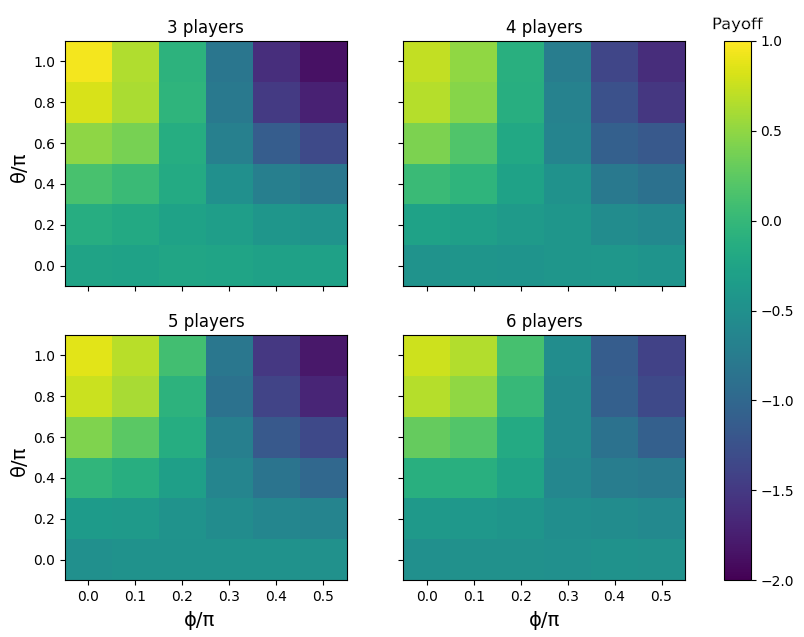}
    \caption{Experimental demonstration of Nash equilibrium for the multiplayer quantum Prisoners' Dilemma for $n=3,4,5$, and $6$ players. The maximum payoff of $1$ for Trader 1 occurs when they choose the parameters $\left( \pi, 0 \right)$. For even numbers of players, the plot exhibits lower contrast compared to odd numbers of players, as the decomposition involves more gates.}
    \label{fig:experimental_results}
\end{figure}

We note in our experiments that while the quantum referee proposed by Du et al. generalizes to $n$ traders, for even values of $n$ greater than 2, the quantum correlated equilibrium arises only when the initial state of every qubit is $\ket{1}$ instead of $\ket{0}$. The Nash equilibrium outcomes for the different numbers of players are exhibited in the heat map of Figure \ref{fig:experimental_results}. For $n>2$, our processor required more gates in the decomposition than for when $n$ is odd. For $n$ players, the number of entangling gates used for each circuit to show Nash equilibrium is on the order of $3n$. 

\subsection{Two-Traders with Mixed Strategies}
\begin{figure}
    \centering
    \includegraphics[width=0.5\textwidth]{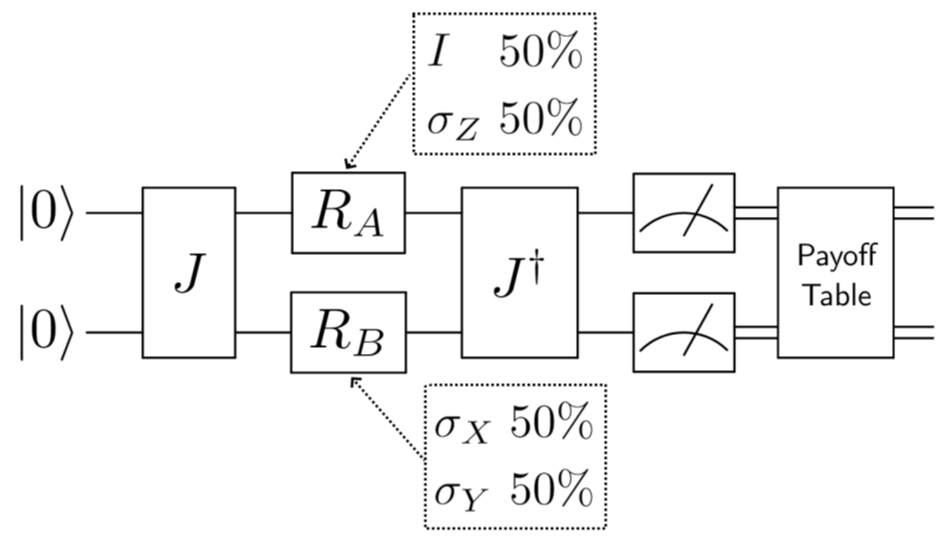}
    \caption{Quantum circuit for Nash equilibrium in two-trader Prisoner's Dilemma (or Chicken) when the game is played probabilistically.}
    \label{fig:randomized_equilibrium}
\end{figure}
To show Nash equilibrium for randomized strategies, we follow a similar method but repeat the process twice since the equilibrium strategy is no longer symmetric.  Explicitly, we fix Trader 2's strategy and allow Trader 1 to vary his quantum strategy parameters $(\alpha,\theta,\gamma)$.  If this is a Nash equilibrium, then Trader 1's payoff is maximized when adopting the parameters according to his mixed strategy.  We then show the same after fixing the other player's strategy. 

\begin{figure}
    \centering
    \includegraphics[width=0.9\textwidth]{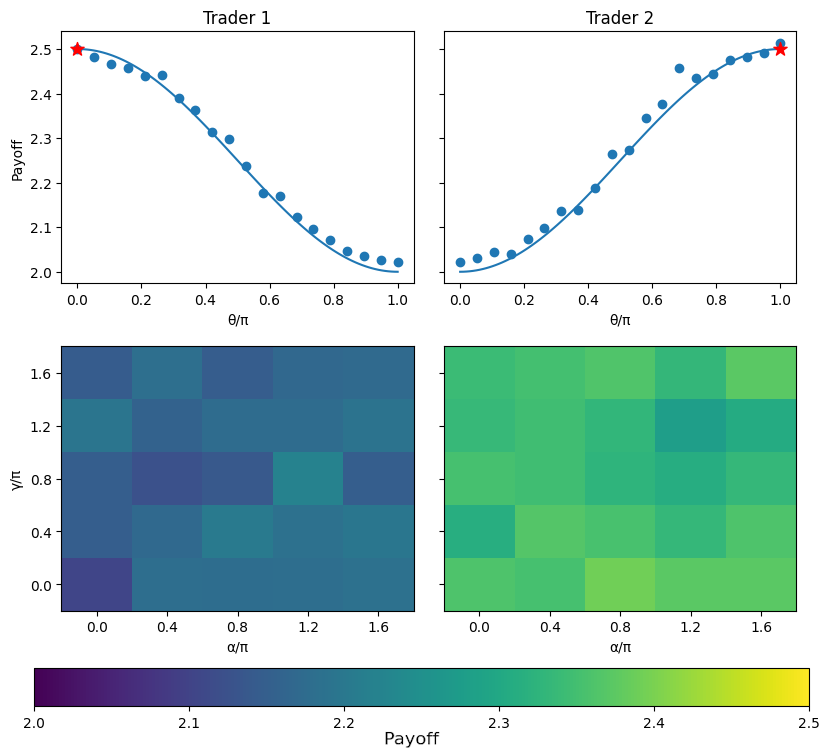}
    \caption{Payoffs of players when deviating from the Nash equilibrium strategy for randomized Prisoners' Dilemma.  Top: Experimentally measured payoffs for different values of $\theta$ are shown.  The solid blue line shows the simulated payoff, and the blue dots show the payoff obtained in experiment.  The red star is the Nash equilibrium strategy.  The maximum payoff is obtained at the Nash equilibrium strategy for both players, confirming the Nash equilibrium.  Bottom: Experimentally measured payoffs for different values of $\alpha$ and $\gamma$ at a fixed value of $\theta$. When one player chooses the Nash equilibrium strategy, the simulated payoffs do not have a dependence on $\alpha$ and $\gamma$, which is confirmed experimentally. }
    \label{fig:randomized_equilibrium1}
\end{figure}

When one player plays their equilibrium strategy, the playoff of both players only depends on the other player's choice of $\theta$ with no dependence on $\alpha$ and $\gamma$.  As such, when testing the Nash equilibrium, we show that the $\theta$ value giving the highest payoff corresponds to that player's equilibrium strategy.  Trader 1 achieves a maximum payoff at $\theta=0$, and Trader 2 achieves a maximum payoff at $\theta = \pi$.  We successfully demonstrated Nash equilibrium for randomized strategies as shown in Figure \ref{fig:randomized_equilibrium1}.

\section{Discussion}

We show how the strategic dynamics of Chicken and Prisoner’s Dilemma manifest in trading and explored their implementation on a quantum trading platform. Our results highlight how a quantum referee can create higher-order correlations and deliver a quantum advantage, achieving superior market Nash equilibria. Additionally, we showcase this Nash equilibrium experimentally on an ion-trap quantum processor for up to six traders.

\begin{figure}
    \centering
     \centering
    \includegraphics[scale=0.55]{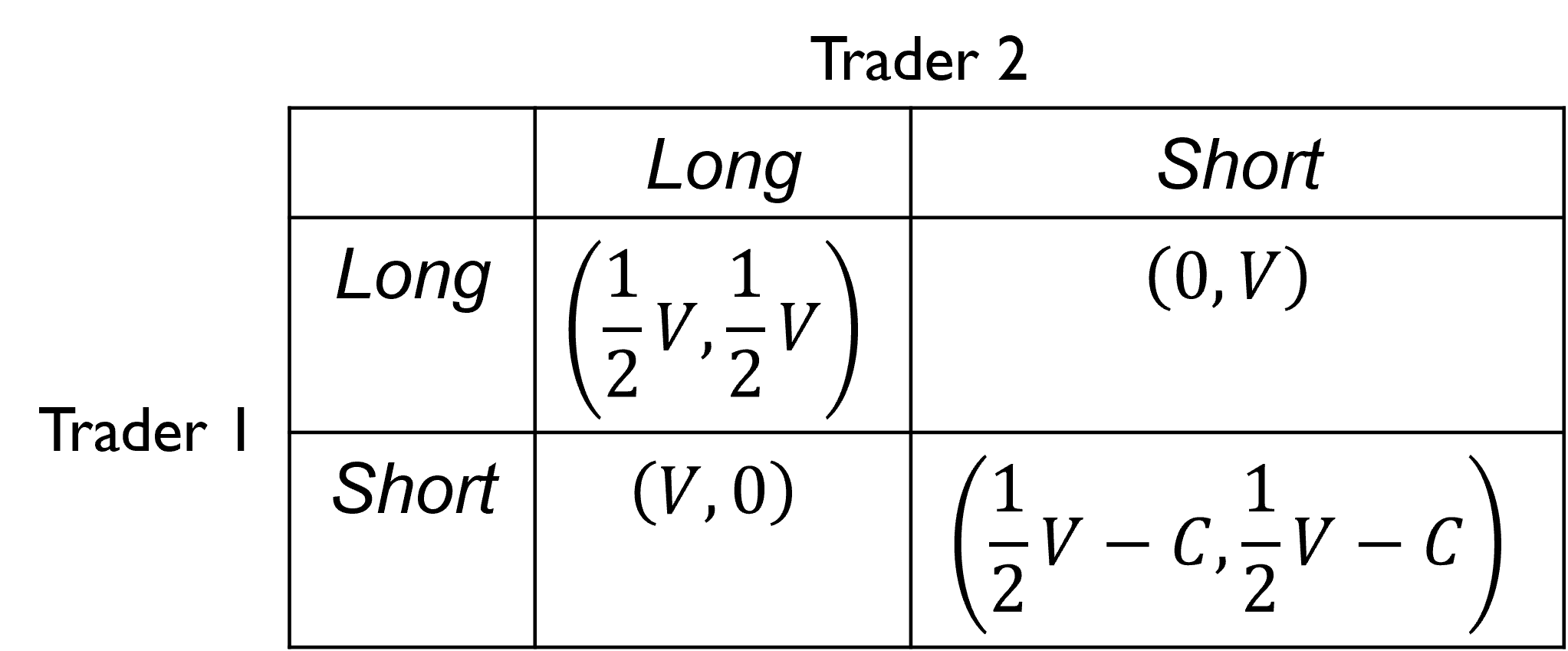} 
    \caption{The class of Hawk-Dove game models for equity trading. The variable $V$ represents the value of the asset being traded, which is split when both traders go long, and the variable $C$ represents the cost of both going short.}
    \label{fig:HD}
\end{figure}

Notably, the quantum referee yields superior Nash equilibria for the entire class of two-player Hawk-Dove games \cite{binmore}. In these games, each player employs two strategies: \textit{Dove}, corresponding to the \textit{Long} position, and \textit{Hawk}, corresponding to the \textit{Short} position, as illustrated in Figure \ref{fig:HD}. The payoff structure ensures that the strategy pairs (\textit{Long, Short}) and (\textit{Short, Long}) result in opposite payoffs to the players. Within this framework, Prisoner’s Dilemma and Chicken emerge as special cases with specific parameter values, namely $V=4$ and $C=1$ or $C=3$, respectively.

The traders/players heed the advice of a quantum referee  even when it is provided probabilistically and realize a better paying Nash equilibrium. In future work, we aim to extend Hawk-Dove games to include additional players and explore more complex game-theoretic scenarios, including mixed quantum strategies. 

We also plan to investigate the implementation of these games on quantum computing platforms, exploring in detail the role of quantum entanglement to leverage advantage in financial applications. For instance, this advantage could create win-win scenarios by achieving superior market equilibrium in green markets. Green markets, such as carbon trading, are designed to facilitate the exchange of environmental pollutants under regulatory restrictions (e.g., ``cap and trade'') with the primary aim of mitigating pollution by sustaining higher prices. However, if these markets adopt a Prisoner's Dilemma dynamic, short-selling could become the dominant strategy, resulting in market volatility and potentially lower prices. Quantum trading technology offers a promising solution to address this issue. 

\section{Acknowledgements}

NML acknowledges support from the U.S. Department of Energy (DoE), Office of Science, National Quantum Information Science Research Centers, Quantum Systems Accelerator (DE-FOA-0002253) and the  National Science Foundation, Software-Tailored Architecture for Quantum Co-Design (STAQ) Award (PHY-2325080). 
The authors thank Alaina Green for experimental support.


\begin{thebibliography}{9}

\bibitem{Dong} Y. Dong et al., {\it A narrative review on quantum finance theory},  International Journal of Quantum Information, Vol. 22, No. 06, 2450016 (2024).

\bibitem{Herman} D. Herman et al., {\it Quantum computing for finance}, Nature Reviews Physics volume 5, pages 450–465 (2023).

\bibitem{Khan} F. Khan et al., {\it Quantum Prisoner’s Dilemma and High Frequency Trading on the Quantum Cloud}, Front. Artif. Intell., 02 November 2021
Sec. AI in Finance, 
Volume 4 - 2021 | https://doi.org/10.3389/frai.2021.769392 

\bibitem{Cruttenden} W. Cruttenden, {\it Shorting America}, available at \url{https://www.sec.gov/comments/4-627/4627-95.pdf}.

\bibitem{binmore} K. Binmore, {\it Playing for Real}, Oxford University Press.

\bibitem{Aumann} R. Aumann, {\it Subjectivity and correlation in randomized strategies}, Journal of Mathematical Economics, Volume 1, Issue 1, March 1974, Pages 67-96.

\bibitem{Bleiler} S. Bleiler, {\it A Formalism for Quantum Games and an Application}, \url{ https://doi.org/10.48550/arXiv.0808.1389}.

\bibitem{EWL1} J. Eisert et al., {\it Quantum Games and Quantum Strategies}, Phys. Rev. Lett. 83, 3077 – Published 11 October, 1999.

\bibitem{Nielson} M. Nielsen, et al., {\it Quantum Computation and Quantum Information}, Cambridge University Press, (2000). 

\bibitem{Clauser} J. Clauser et al., {\it Proposed Experiment to Test Local Hidden-Variable Theories}, Phys. Rev. Lett. 23, 880 – Published 13 October, 1969 Erratum Phys. Rev. Lett. 24, 549 (1970)

\bibitem{Solmeyer} N. Solmeyer et al., {\it Demonstration of a Bayesian quantum game on
an ion-trap quantum computer}, 2018 Quantum Sci. Technol. 3 045002.

\bibitem{Chap} J. Chappell et al., {\it N-Player Quantum Games in an EPR Setting}. PLoS ONE 7(5): e36404. \url{https://doi.org/10.1371/journal.pone.0036404}.

\bibitem{EWL2} J. Eisert et al., {\it Quantum games}, Journal of Modern Optics, 47(14–15), 2543–2556. \url{https://doi.org/10.1080/09500340008232180}.

\bibitem{Magli} A. C. Magli et al., {\it The Tragedy of the Commons as a Prisoner’s Dilemma. Its Relevance for Sustainability Games}, Sustainability 2021, 13, 8125. \url{https://doi.org/10.3390/su13158125}. 

\bibitem{Du} J. Du et al., {\it Entanglement Enhanced Multiplayer Quantum Games}, Physics Letters A, Volume 302, Issues 5–6, 30 September 2002, Pages 229-233.

\bibitem{Dax} D. Koh et al., {\it Quantum Volunteer's Dilemma}. Preprint available at  arXiv:2409.05708 [quant-ph].

\bibitem{Ahmed} A. Ahmed, {\it Quaternions, Octonions, and the Quantization of Games}, doctoral dissertation, (2009). \url{https://pdxscholar.library.pdx.edu/open_access_etds/5944/}.

\bibitem{Debnath} S. Debnath et al., {\it Demonstration of a small programmable quantum computer with atomic qubits}, Nature, Volume 536, Number 7624, 01 August 2016, Pages 63-66.

\bibitem{Olmschenk} S. Olmschenk et al., {\it Manipulation and detection of a trapped Yb+ hyperfine qubit}, Phys. Rev. A, Volume 76, Issue 5, 19 November 2007.
\url{https://journals.aps.org/pra/abstract/10.1103/PhysRevA.76.052314}

\bibitem{MS} A. Sørensen and K. Mølmer, {\it Quantum Computation with Ions in Thermal Motion}, Phys. Rev. Lett. 82, 1971, 1 March 1999
\url{https://journals.aps.org/prl/abstract/10.1103/PhysRevLett.82.1971}

\end{thebibliography}
\end{document}